\documentclass{iauc}
\usepackage{graphicx}
\pubyear{2005}
\volume{199}
\pagerange{1--}
\jname{Probing Galaxies through Quasar Absorption Lines}
\editors{P. R. Williams, C. Shu, and B. M\'{e}nard, eds.}

\newcommand{\kms}{km~s$^{-1}\,$}
\newcommand{\ms}{m~s$^{-1}\,$}

\newcommand{\daa}{$\Delta\alpha/\alpha\,$}

\newcommand{\FeII}{Fe\,{\sc ii}~}

\newcommand{\zabs}{$z_{\rm abs}$}

\title[Cosmological variability of  $\alpha$ ] 
{VLT/UVES shows no cosmological variability of  $\alpha$ }

\author[P. Molaro, M. Centurion, S. D'Odorico, S. Levshakov]   
{Paolo Molaro$^1$, 
Miriam Centurion$^1$, 
Sandro D'Odorico$^2$,  \and \break Sergei Levshakov$^3$}
\affiliation{$^1$Osservatorio Astronomico di Trieste-INAF, Via G.B. Tiepolo 11 I-34131, Trieste, Italy \break 
\\[\affilskip]
$^2$ESO, Karl-Schwarzschild-Strasse 2 D-85748 Garching bei Munchen, Germany \break 
\\[\affilskip]
$^3$ Ioffe Physico-Technical Institute, St. Petersburg,
Russia}

\begin{document}

\maketitle

\begin{abstract}
The cosmological
variability of $\alpha$ is probed from   individual observations of 
pairs of \FeII lines. 
This procedure  allows a better control of the systematics and  avoids  the influence of the spectral shifts due to
 ionization inhomogeneities in the absorbers 
and/or non-zero offsets between different exposures. 
Applied to the \FeII lines of the metal absorption systems  at 
\zabs = 1.839 in Q1101--264 and at  \zabs = 1.15 in HE0515--4414 observed by means of 
UVES at the ESO-VLT, it
provides \daa = $(0.4\pm1.5_{\rm stat})\times10^{-6}$.
The result is  shifted with respect to  the Keck/HIRES mean
\daa = $(-5.7\pm1.1_{\rm stat})\times10^{-6}$ 
(Murphy et al. 2004) at a high confidence level (95\%).
Full details of this work  are given in Levshakov et al (2005)  

\end{abstract}

\firstsection 
\section{Introduction}

The Sommerfeld fine-structure constant  
$\alpha \equiv e^2/\hbar c$ describes 
electromagnetic and optical properties of atoms and  is the most suitable 
constant for time variation tests   in the lab  or for
astronomical observations  
(for a review  see Uzan 2003). The value  is  
$\alpha = 1/137.035\,999\,76(50)$ (Mohr \& Tailor 2000) and
laboratory experiments constrain its variability as $d$ln[$\alpha(t)]/dt = (-0.9\pm2.9_{\rm stat})\times10^{-15}$ yr$^{-1}$
(Fischer et al. 2004). This limit corresponds to  
 $|\Delta \alpha/\alpha| \equiv |(\alpha_z - \alpha)/\alpha| < 3.8\times10^{-5}$
 for $t \sim 10^{10}$ yr\,  and   $\alpha_z$ varying linearly with time, which may be not the case. Analysis of radioactive decay rates of   nuclei
 in meteorites set  
$\Delta\alpha/\alpha = (+8\pm8_{\rm stat})\times10^{-7}$ (Olive et al. 2004) while
a recent analysis of the isotopic abundances in the Oklo samples 
($\Delta t \sim 2\times10^9$ yr) suggests  
\daa $\geq 4.5\times10^{-8}$ (Lamoreaux \& Torgerson 2004). 
Shifts in the alkali doublets  provide
$\Delta\alpha/\alpha = (-0.5\pm1.5_{\rm stat})\times10^{-5}$ 
(Murphy et al. 2001) but the Many-Multiplet  method applied to 143  absorption systems 
in the Keck/HIRES
spectra of quasars indicates
a decrease of $\alpha$:
$\Delta\alpha/\alpha = (-5.7\pm1.1_{\rm stat})\times10^{-6}$ in the redshift range
$0.2 < z < 4.2$ (Murphy et al. 2004, MFWDPW).
However,   VLT/UVES observations of 23 absorption systems 
($0.4 \leq z \leq 2.3$) toward 18 QSOs show no variability:
$\Delta\alpha/\alpha = (-0.6\pm0.6_{\rm stat})\times10^{-6}$ (Chand et al 2004). A first result of the new approach  presented here applied to the system  
at \zabs = 1.15  towards 
HE0515--4414 ($B = 15.0$) was
$\Delta\alpha/\alpha = (-0.4\pm1.9_{\rm stat} \pm2.7_{\rm sys}) \times 10^{-6}$ 
(Quast, Reimers \& Levshakov 2004), also showing no variability of $\alpha$.
  
\section{The concept}

  We  use a modified Many-Multiplet procedure to calculate \daa\, {\it directly} from the 
differences between the wavelengths of a pair of \FeII transitions
observed in the individual exposures.   We call it SIDAM
 {\it Single Ion Differential $\alpha$ Measurement}. Full details of the method and of the analysis presented here are given in Levshakov et al (2005).
\begin{itemize}
\item Why \FeII? 

 Dealing with only one heavy element  we reduce:
({\it i}) The influence of unknown isotopic ratio. For  \FeII  this is   less pronounced than that
for MgII because  iron is heavier and its isotope structure is more compact. The relative abundance of the leading isotope $^{56}$Fe is 
higher (terrestrial isotope ratios are
$^{54}$Fe: $^{56}$Fe: $^{57}$Fe: $^{58}$Fe = 5.8 : 91.8 : 2.1 : 0.3, and 
$^{24}$Mg: $^{25}$Mg: $^{26}$Mg = 79 : 10 : 11).
({\it ii}) The effects of possible inhomogeneous ionization structure within the absorber.
({\it iii})  For \FeII the relativistic correction
to the changes in $\alpha$  ($q$ coefficients)  are rather large. 

\item  Why single exposures? 

Being applied to the lines from the {\it same}
exposure, this method does not depend
on the {\it unknown offsets of the wavelength scale}.
There are three main sources of potential  
systematic errors in the absolute velocity scale:
($i$) temperature, 
 ($ii$) air pressure instability, 
 ($iii$)
mechanical instabilities of unknown origin. 
For instance, a change of 1 millibar, or a change of
0.3$^\circ$C,  corresponds to   an error in radial velocities of $\sim 50$ \ms,
or 
$\Delta\alpha/\alpha \sim 1.7\times10^{-6}$.
\end{itemize}

\section{Observations and data reduction}

Five high resolution UVES spectra
of Q1101--264 ($V = 16.02$) with a \zabs = 1.839 system were obtained  with a dichroic filter during the  Science Verification programme on Feb. 2000.  
The  spectral resolution as measured from the ThAr emission lines are of
FWHM $\simeq 6.0$ \kms\, in the blue ($\lambda \sim 4570$~\AA), 
and of $\simeq 5.4$ \kms\, in the red ($\lambda \sim 7380$~\AA). 
  We modified the UVES pipeline  to calibrate  the echelle spectra without rebinning
in wavelength to preserve  the original pixel size: 50 m\AA\, (3.3 \kms/pix) for the blue and
55 m\AA\, (2.2 \kms/pix) for the red.
Typical rms of the wavelength calibration are of  $\approx 1$~m\AA\, ($\approx 60$ \ms).
 We worked with individual exposures, since co-adding  them     cancels out  original non-zero
offsets   and   perturbs  line centroids.
 All spectra are corrected to  heliocentric and vacuum scale
 but  they are {\it not combined} and {\it not resampled}.

 \begin{figure}
 \includegraphics[height=3in,width=4in,angle=0]{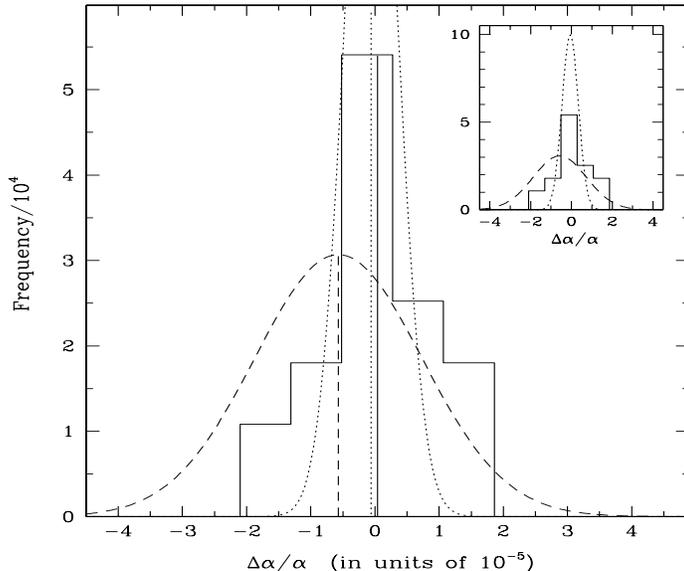}
  \caption{The histogram shows the distribution of \daa 
from the \FeII
systems in  Q1101--264 and HE0515--4414. 
The dashed curves show the distributions  from MFWDPW and CSPA. Note
the discrepancy between  Keck/HIRES and VLT/UVES 
sample mean  values.
 }\label{fig:fg6}
\end{figure}
 \section{Discussion}
Measurements for Q1101--264  are  combined with the  \FeII sample from
the \zabs = 1.15 system toward HE0515--4414 (QRL) providing a  total of
35 \daa values.
The normalized distribution 
of the resulting 35 \daa values is plotted with  histogram  in Fig.~1. In the figure the  results of MFWDPW and CSPA   are also shown
by the dashed and dotted curves, respectively, assuming that the measured
\daa are normally distributed with the sample means and standard deviations
 taken from  the original  papers. 
The vertical lines in this figure
mark the centers of the corresponding distributions.
Our result is  shown by the histogram and has the sample mean:

\bigskip

\centerline{
$\langle$\daa$\rangle$ = $(0.4\pm1.5_{\rm stat})\times10^{-6}$.}

\bigskip\noindent 

The scatter of \daa in the Keck sample is
about 2 times the $\sigma_{\rm rms}$ value of our combined \FeII sample, but the sample means of CSPA and our \FeII ensemble are in good agreement. 
The results presented in Fig.1 show that
$\langle$\daa$\rangle_{\rm Keck/HIRES} \neq$
$\langle$\daa$\rangle_{\rm VLT/UVES}$. Our result, and also the one of  CSPA, differ from that of MFWDPW at  95\% ~ significance level according to 
the $t$-test.  

We  also note that    standard deviation in the CSPA sample is exceptionally 
 small  somewhat in disagreement with their  
  wavelength calibration  accuracy. This  is checked through the relative velocity
shifts, $\Delta v$, between the \FeII $\lambda2344$ and 
$\lambda2600$ lines. The mean $\langle \Delta v \rangle$  
shows a dispersion of
$\sigma_{\Delta v} \simeq 0.4$ \kms, equivalent to $\sigma_{\Delta\alpha/\alpha} \sim 2\times10^{-5}$, which is not consistent with an error of the mean 
$\sigma_{\langle \Delta \alpha/\alpha \rangle}
\sim 0.6\times10^{-6}$ as CSPA  claimed. 

Although the result presented here is consistent with no evolution of  $\alpha$,  we emphasize that
 accurate individual measurements of $\langle$\daa$\rangle$ are required if $\alpha$ has an  oscillating behaviour as suggested by Fuji (2005).

\end{document}